\documentclass[aps,pre,twocolumn,groupedaddress]{revtex4}

\usepackage{graphicx}
\usepackage{dcolumn}
\usepackage{amssymb}
\usepackage{bm}

\begin{document}

%\preprint{}

%Title of paper
\title{ 
Line tension of branching junctions of bilayer membranes}

\author{Hiroshi Noguchi}
\email[]{noguchi@issp.u-tokyo.ac.jp}
\affiliation{
Institute for Solid State Physics, University of Tokyo,
 Kashiwa, Chiba 277-8581, Japan}
%\homepage[]{Your web page}
%\altaffiliation{}

%\date{\today}

\begin{abstract}
Branching of bilayer membranes appear
in the inverted hexagonal phase as well as
in metastable states of the lamellar phase such as membrane fusion intermediates.
A method for estimating the line tension of the branching junction is 
proposed for molecular simulations.
The line tension is calculated from the pressure tensor of
 equiangularly branched membranes.
The simulation results agree very well with the theoretical prediction of
Hamm and Kozlov's tilt model.
The transition between the lamellar and inverted hexagonal phases is also 
investigated using the tilt model.
\end{abstract}

\maketitle

\section{Introduction}

Amphiphilic molecules, such as lipids and detergents,
 self-assemble into various structures:
spherical or cylindrical micelles; lamella; and
hexagonal, cubic, and sponge structures \cite{isra11,lipo95,grun85,grun89}.
The bilayer membrane in the lamellar phase is the basic structure
of the plasma membrane and the intracellular compartments of living cells.
It is known that biological membranes contain a substantial amount
of lipids that can adopt the inverted hexagonal $H_{II}$ phase 
upon isolation \cite{verk84,epan98,hafe01}. 
Thus, it has been speculated that non-bilayer structures play some roles
in biological functions such as
 membrane fusion and other membrane contact phenomena such as tight junctions.

Several mechanisms of membrane fusion have been proposed \cite{jahn02,cher08,niko11,mark11,mull11}.
Among these, the stalk model is widely accepted.
According to this model, two vesicles form a metastable structure, the stalk intermediate,
where the outer monolayers (leaflets) are connected with a cylindrical (stalk) structure.
Next, radial expansion of the stalk \cite{cher08,sieg93} 
or a pore opening on the side of the stalk  \cite{mull11,nogu01b}
leads to either complete fusion or the formation of a hemifusion diaphragm (HD) intermediate.
In the HD, the inner monolayers of two vesicles form
a bilayer membrane, which connects with two bilayer membranes of the vesicles 
via a branching junction. 
Many experiments support that fusion proceeds through a metastable intermediate state,
which is considered to be the stalk or HD intermediate.
Recently, molecular simulations of membrane fusion and fission have been reported by several groups 
\cite{mark11,mull11,nogu01b,nogu02a,nogu02b,nogu02c,muel02,marr03,li05,smei06,knec07,gao08}.
In most of these simulations, the stalk intermediate is formed first.
However, the pathway of the stalk to complete fusion
is not unique and is dependent on the simulation conditions.
Thus far, it has not been determined
which membrane properties and external conditions
determine the fusion pathways.
It is important to estimate the free energy of each fusion intermediate
to understand the fusion pathways.

Since the transition between lamellar and $H_{II}$ phases requires 
the same type of topological changes as membrane fusion,
it is considered to proceed through similar pathways \cite{sieg93,li00}.
Recently, the phase of a hexagonally ordered stalk structure 
has been observed between lamellar and $H_{II}$ phases \cite{yang02}.
Here, we consider the $H_{II}$ structure with a long cell length.
The bilayer membranes are connected with hexagonally arranged branching junctions.
When the branching junctions of bilayers are isolated, {\it i.e.}, sufficiently far from other junctions,
the free energy of a junction is proportional to the junction length.
Thus, it can be estimated as a line tension.
In this paper, we propose a method for estimating this line tension for molecular simulations.
This tension quantifies the stability of the HD fusion intermediate.

To simulate lipid membranes on long and large scales,
various coarse-grained molecular models have been proposed (see review articles \cite{muel06,vent06,nogu09,marr09}).
Recently, the bottom-up approaches have been intensively investigated to construct coarse-grained molecular models, 
in which potential parameters are tuned from atomistic simulations \cite{marr04,izve05,arkh08,shin08,wang10}.
We, on the other hand, choose an opposite top-down approach to construct the coarse-grained molecular models, 
in which potentials are based on the continuum theory.
Here, we employ a spin lipid molecular model \cite{nogu11}, which is one of the solvent-free molecular lipid models \cite{nogu09}.
In this model, the membrane properties (bending rigidity, line tension of membrane edge,
area compression modulus, lateral diffusion coefficient, and flip-flop rate) can be varied over wide ranges.

In Sec. \ref{sec:method}, the simulation method and the membrane properties of the simulation model 
are described.
In Secs. \ref{sec:results} and \ref{sec:tilt}, 
the line tension is estimated by the simulation and the tilt model, respectively.
The line tension is calculated from the pressure tensor of hexagonally arranged membrane branches.
Previously, the line tension of the membrane edge has been estimated from the pressure tensor of a membrane strip \cite{tolp04,wang10,nogu11}.
For branching junctions the membrane can hold finite surface tension, unlike the strip,
hence, the membrane area should be carefully treated.
The tilt model proposed by Hamm and Kozlov \cite{hamm98,hamm00} reproduces
the simulation results very well.
In Sec. \ref{sec:dynamic}, the rupture dynamics of the branches and the HD intermediates of the vesicles
are shown. Finally, a summary is given in Sec. \ref{sec:sum}.

\section{Simulation Model and Method}
\label{sec:method}

\subsection{Membrane Model}

A spin lipid molecular model \cite{nogu11} is used to simulate bilayer membranes.
Each ($i$-th) molecule has a spherical particle with an orientation vector ${\bf u}_i$, 
which represents the direction from the hydrophobic to the hydrophilic part ($|{\bf u}_i|=1$).
There are two points of interaction in the molecule:
the center of a sphere ${\bf r}^{\rm s}_i$ and a hydrophilic point ${\bf r}^{\rm e}_i={\bf r}^{\rm s}_i+{\bf u}_i \sigma$.
The molecules interact with each other via the potential,
\begin{eqnarray}
\frac{U}{k_{\rm B}T} &=\ \ & \hspace{1cm} \sum_{i<j} U_{\rm {rep}}(r_{ij}^{\rm s}) \label{eq:U_all}
               +\varepsilon \sum_{i} U_{\rm {att}}(\rho_i)  \\ \nonumber
&\ \ +& \ \ \frac{k_{\rm{tilt}}}{2} \sum_{i<j} \bigg[ 
( {\bf u}_{i}\cdot \hat{\bf r}^{\rm s}_{ij})^2
 + ({\bf u}_{j}\cdot \hat{\bf r}^{\rm s}_{ij})^2  \bigg] w_{\rm {cv}}(r^{\rm e}_{ij}) \\ \nonumber
&\ \ +&  \frac{k_{\rm {bend}}}{2} \sum_{i<j}  \bigg({\bf u}_{i} - {\bf u}_{j} - C_{\rm {bd}} \hat{\bf r}^{\rm s}_{ij} \bigg)^2 w_{\rm {cv}}(r^{\rm e}_{ij}),
\end{eqnarray} 
where ${\bf r}_{ij}={\bf r}_{i}-{\bf r}_j$, $r_{ij}=|{\bf r}_{ij}|$,
 $\hat{\bf r}_{ij}={\bf r}_{ij}/r_{ij}$, and $k_{\rm B}T$ is the thermal energy.
The molecules have an excluded volume with a diameter $\sigma$ via the repulsive potential,
$U_{\rm {rep}}(r)=\exp[-20(r/\sigma-1)]$,
with a cutoff at $r=2.4\sigma$.

The second term in Eq. (\ref{eq:U_all}) represents the attractive interaction between the molecules.
An attractive multibody potential $U_{\rm {att}}(\rho_i)$ is 
employed to mimic the ``hydrophobic'' interaction.
This potential allows the formation of  the fluid membrane over wide parameter ranges.
Similar potentials have been applied in other membrane models \cite{nogu01a,nogu06,fara09}
 and a coarse-grained protein model \cite{taka99}.
The potential $U_{\rm {att}}(\rho_i)$ is given by
\begin{eqnarray} \label{eq:U_att}
U_{\rm {att}}(\rho_i) = 0.25\ln[1+\exp\{-4(\rho_i-\rho^*)\}]- C,
\end{eqnarray} 
with $C= 0.25\ln\{1+\exp(4\rho^*)\}$.
The local particle density $\rho_i$ is approximately the number of
particles ${\bf r}^{\rm s}_i$  in the sphere with radius $r_{\rm {att}}$.
\begin{equation}
\rho_i= \sum_{j \ne i} f_{\rm {cut}}(r^{\rm s}_{ij}), 
\label{eq:wrho}
\end{equation} 
where $f_{\rm {cut}}(r)$ is a $C^{\infty}$ cutoff function,
\begin{equation} \label{eq:cutoff}
f_{\rm {cut}}(r)=\left\{ 
\begin{array}{ll}
\exp\{A(1+\frac{1}{(r/r_{\rm {cut}})^n -1})\}
& (r < r_{\rm {cut}}) \\
0  & (r \ge r_{\rm {cut}}) 
\end{array}
\right.
\end{equation}
with $n=6$, $A=\ln(2) \{(r_{\rm {cut}}/r_{\rm {att}})^n-1\}$,
$r_{\rm {att}}= 1.9\sigma$  $(f_{\rm {cut}}(r_{\rm {att}})=0.5)$, 
and the cutoff radius $r_{\rm {cut}}=2.4\sigma$.
The density $\rho^*$ in $U_{\rm {att}}(\rho_i)$ is the characteristic density.
For $\rho_i < \rho^*-1$,
$U_{\rm {att}}(\rho_i)$ acts as a pairwise attractive potential 
while
it approaches a constant value for $\rho_i > \rho^*+1$.

The third and fourth terms in Eq.~(\ref{eq:U_all}) are
discretized versions of the
tilt and bending potentials of the tilt model \cite{hamm98,hamm00}, respectively.
A smoothly truncated Gaussian function~\cite{nogu06} 
is employed as the weight function 
\begin{equation} \label{eq:wcv}
w_{\rm {cv}}(r)=\left\{ 
\begin{array}{ll}
\exp (\frac{(r/r_{\rm {ga}})^2}{(r/r_{\rm {cc}})^n -1})
& (r < r_{\rm {cc}}) \\
0  & (r \ge r_{\rm {cc}}) 
\end{array}
\right.
\end{equation}
with  $n=4$, $r_{\rm {ga}}=1.5\sigma$, and $r_{\rm {cc}}=3\sigma$.
All orders of derivatives of $f_{\rm {cut}}(r)$ and $w_{\rm {mls}}(r)$ 
are continuous at the cutoff radii.
The weight is a function of $r^{\rm e}_{ij}$
to avoid the interaction between the molecules in the opposite monolayers of the bilayer.
If  $r^{\rm s}_{ij}$ is employed instead, a single-layer membrane is formed \cite{shiba11}.

\subsection{Simulation Method}

The bilayer membrane is simulated in the $NVT$ ensemble (constant
number of molecules $N$, volume $V$, and temperature $T$)
and Brownian dynamics (molecular dynamics with Langevin thermostat) is employed.
The motion of the center of the mass 
${\bf r}^{\rm G}_{i}=({\bf r}^{\rm s}_{i}+{\bf r}^{\rm e}_{i})/2 $ and 
the orientation ${\bf u}_{i}$ are given by underdamped Langevin equations:
\begin{eqnarray}
  \frac{d {\bf r}^{\rm G}_{i}}{dt} &=& {\bf v}^{\rm G}_{i}, \ \  \frac{d {\bf u}_{i}}{dt} = {\boldsymbol \omega}_{i}, \\
m \frac{d {\bf v}^{\rm G}_{i}}{dt} &=&
 - \zeta_{\rm G} {\bf v}^{\rm G}_{i} + {\bf g}^{\rm G}_{i}(t)
 + {\bf f}^{\rm G}_i, \\
I \frac{d {\boldsymbol \omega}_{i}}{dt} &=&
 - \zeta_{\rm r} {\boldsymbol \omega}_i + ({\bf g}^{\rm r}_{i}(t)
 + {\bf f}^{\rm r}_i)^{\perp} + \lambda_{\rm L} {\bf u}_{i},
\end{eqnarray}
where $m$ and $I$ are the mass and the moment of inertia of the molecule, respectively.
The forces are given by ${\bf f}^{\rm G}_i= - \partial U/\partial {\bf r}^{\rm G}_{i}$
and ${\bf f}^{\rm r}_i= - \partial U/\partial {\bf u}_{i}$ with 
the perpendicular component ${\bf a}^{\perp} ={\bf a}- ({\bf a}\cdot{\bf u}_{i}) {\bf u}_{i}$
and a Lagrange multiplier $\lambda_{\rm L}$ to keep ${\bf u}_{i}^2=1$.
According to  the fluctuation-dissipation theorem,
the friction coefficients $\zeta_{\rm G}$ and $\zeta_{\rm r}$ and 
the Gaussian white noises ${\bf g}^{\rm G}_{i}(t)$ and ${\bf g}^{\rm r}_{i}(t)$
obey the following relations:
the average $\langle g^{\beta_1}_{i,\alpha_1}(t) \rangle  = 0$ and the variance
$\langle g^{\beta_1}_{i,\alpha_1}(t_1) g^{\beta_2}_{j,\alpha_2}(t_2)\rangle  =  
         2 k_{\rm B}T \zeta_{\beta_1} \delta _{ij} \delta _{\alpha_1 \alpha_2} \delta _{\beta_1 \beta_2} \delta(t_1-t_2)$,
where $\alpha_1, \alpha_2 \in \{x,y,z\}$ and  $\beta_1, \beta_2 \in \{{\rm G, r}\}$.
The Langevin equations are integrated by the leapfrog algorithm \cite{alle87,nogu11}.

The results are displayed with a length unit of $\sigma$, an energy unit of $k_{\rm B}T$, and
 a time unit of $\tau_0=\zeta_{\rm G}\sigma^2/k_{\rm B}T$.
 $m= \zeta_{\rm G}\tau_0$, $I=\zeta_{\rm r}\tau_0$, $\zeta_{\rm r}=\zeta_{\rm G}\sigma^2$,
 $\Delta t=0.005\tau_0$, and $\rho^*=14$ are used.
The error bars of the data are estimated 
from the standard deviations of three or six independent runs.

\subsection{Membrane Properties}
\label{sec:mempro}

\begin{figure}
\includegraphics{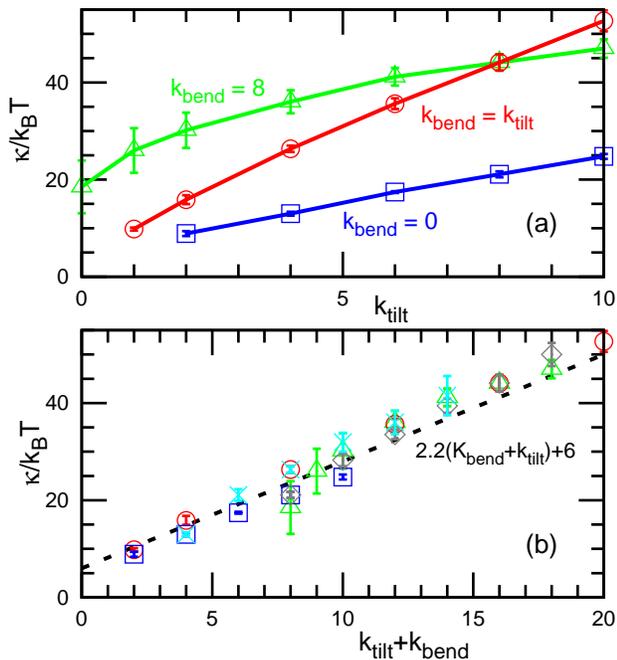}
\caption{\label{fig:kappa}
Bending rigidity $\kappa$ dependence on (a) $k_{\rm {tilt}}$
and (b) $k_{\rm {tilt}}+k_{\rm {bend}}$ for  $\varepsilon=2$ and $C_{\rm {bd}}=0$.
The lines with squares, circles, and triangles represent data 
for $k_{\rm {bend}}=0$, $k_{\rm {bend}}=k_{\rm {tilt}}$, and $k_{\rm {bend}}=8$, respectively.
The crosses and diamonds represent data for $k_{\rm {tilt}}=4$ and $8$,
 respectively, for
several values of $k_{\rm {bend}}$.
The dashed line shows $\kappa/k_{\rm B}T=2.2(k_{\rm {tilt}}+k_{\rm {bend}})+6$.
}
\end{figure}

The potential-parameter dependence of
the membrane properties (bending rigidity $\kappa$, line tension $\Gamma$ 
of membrane edge,
area compression modulus, lateral diffusion coefficient, and flip-flop rate) 
are investigated in detail in our previous paper~\cite{nogu11}.
In this model, these properties can be varied over wide ranges.
The line tension $\Gamma$ of the membrane edge can be controlled by  varying
$\varepsilon$ and $C_{\rm {bd}}$.
The membrane has a wide range of fluid phases, and
the fluid-gel transition point can be controlled by  $\rho^*$.
The area compression modulus $K_{\rm A}$ can be varied by $k_{\rm {tilt}}$.
The flip-flop rate can be varied by $k_{\rm {tilt}}$ and $k_{\rm {bend}}$.

\begin{figure}
\includegraphics{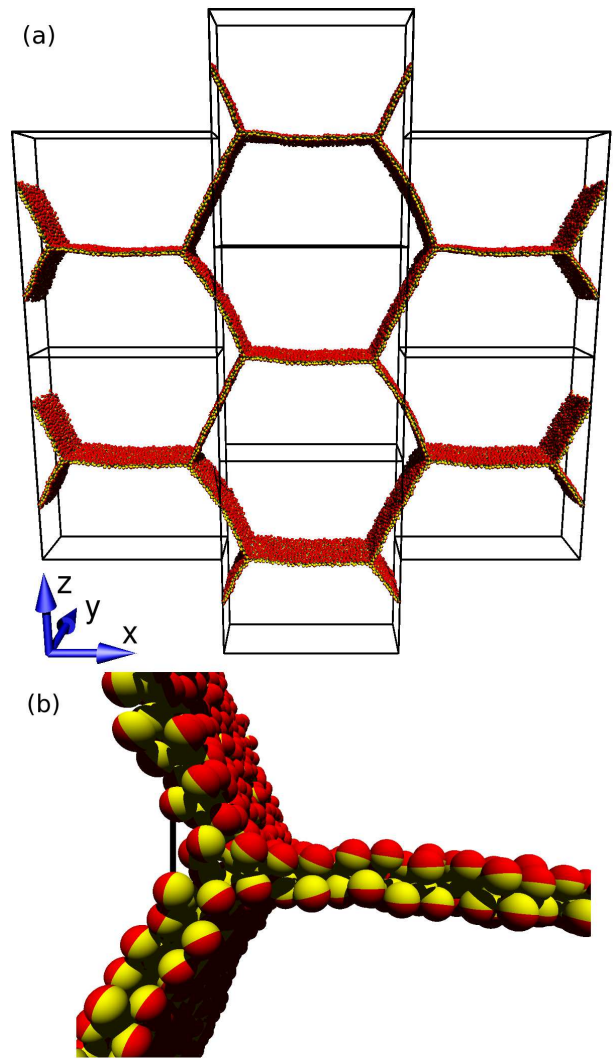}
\caption{\label{fig:hex_snap}
Snapshot of the bilayer membranes with hexagonally arranged branches
for $N=3072$, $L_{x}=\sqrt{3}L_z/2=56\sigma$, $L_{y}=16.94\sigma$,  $\varepsilon=2$, 
$k_{\rm {bend}}=8$, $k_{\rm {tilt}}=8$, and $C_{\rm {bd}}=0$.
(a) Periodic images.
(b) Magnified snapshot of the branching junction.
The red and yellow himispheres represent the hydrophilic and hydrophobic parts of the molecules,
respectively.
}
\end{figure}

The bending rigidity $\kappa$
is linearly dependent on $k_{\rm {tilt}}$ and $k_{\rm {bend}}$.
The spectra of undulation
modes $\langle |h(q)|^2\rangle =k_{\rm B}T/\kappa q^4$ 
are widely used to estimate $\kappa$ for tensionless planar membranes.
In Ref.~\cite{nogu11},
we use a least-squares fit to $1/\langle |h(q)|^2\rangle 
= \kappa q^4/k_{\rm B}T$ for low frequency modes $q<q_{\rm {cut}}$ with $(q_{\rm {cut}}/\pi)^2=0.015$.
More recently, we have found that the extrapolation to $q_{\rm {cut}} \to 0$
yields a better $\kappa$ estimation \cite{shiba11}.
Re-estimated values of $\kappa$ are shown in Fig. \ref{fig:kappa}
for the tensionless membrane
using this  extrapolation method.
Slightly higher values are obtained
and the slope $2$ in Ref.~\cite{nogu11} becomes $2.2$
as $\kappa/k_{\rm B}T=2.2(k_{\rm {tilt}}+ k_{\rm {bend}})+6$
for $\varepsilon=2$ [see Fig. \ref{fig:kappa}(b)].

The bending elasticity generated by the bending and tilt potentials can be
derived from the continuum theory \cite{safr94,helf73} as discussed in Refs. \cite{nogu11,shiba11}.
When the orientation vectors ${\bf u}_{i}$ are equal to the normal
 vectors ${\bf n}$ of the membrane without tilt deformation,
the bending and tilt energies of the monolayer membrane are given by
\begin{eqnarray}
U_{\rm {cv}} &=& 
 \int dA\  \frac{\kappa^*_{\rm {bend}}+\kappa^*_{\rm {tilt}}}{2} 
(C_1+C_2-C_0)^2  \nonumber \\ \label{eq:cv1} 
& & \hspace{0.9cm} - (\kappa^*_{\rm {bend}}+\kappa^*_{\rm {tilt}})C_1 C_2
\end{eqnarray}
in the continuum limit, where
$C_1$ and $C_2$ are the two principal curvatures of
the monolayer membrane.
The bending rigidity and the saddle-splay modulus of the monolayer are given by
 $\kappa_{\rm {mo}}=\kappa^*_{\rm {bend}}+\kappa^*_{\rm {tilt}}$
and  $\bar{\kappa}_{\rm {mo}}= -\kappa$, respectively.
The spontaneous curvature is given by 
$C_0= \{\kappa^*_{\rm {bend}}/(\kappa^*_{\rm {bend}}+\kappa^*_{\rm {tilt}})\}C_{\rm {bd}}/\bar{r}_{\rm {nb}}$,
where $\bar{r}_{\rm {nb}}$ is the nearest-neighbor distance.
By assuming a hexagonal packing of the molecules,
the bending rigidities generated by  the bending and tilt potentials are estimated as 
$\kappa^*_{\rm {bend}}/k_{\rm B}T= \sqrt{3} k_{\rm {bend}} w_{\rm {cv}}(\bar{r}_{\rm {nb}})$ and
$\kappa^*_{\rm {tilt}}/k_{\rm B}T= \sqrt{3} k_{\rm {tilt}} w_{\rm {cv}}(\bar{r}_{\rm {nb}})/2$, respectively.
Thus, the bending rigidity $\kappa$ of the bilayer membrane is estimated as 
$\kappa=2\kappa_{\rm {mo}} \simeq (2.1 k_{\rm {bend}} + 1.1 k_{\rm {tilt}})k_{\rm B}T$ 
from $\bar{r}_{\rm {nb}} \simeq 1.05\sigma$ and $w_{\rm {cv}}(1.05\sigma) = 0.61$.
This estimation of $\kappa$ shows quantitative agreement with the simulation results for $k_{\rm {bend}}$,
while the prefactor of $k_{\rm {tilt}}$ is half that of the numerical estimation.
Deviation from a hexagonal lattice
and tilt deformation likely change the prefactor of $k_{\rm {tilt}}$.
Using $\kappa^{**}_{\rm {tilt}}=2\kappa^*_{\rm {tilt}}$,
the spontaneous curvature 
$C_0  \simeq \{k_{\rm {bend}}/(k_{\rm {bend}}+k_{\rm {tilt}})\}C_{\rm {bd}}/\sigma$  is obtained,
which agrees well with $C_0$ calculated from a membrane tube and strip \cite{shiba11}.
Positive spontaneous curvature indicates that the hydrophilic head is larger than the hydrophobic tail of amphiphilic molecules.

\begin{figure}
\includegraphics{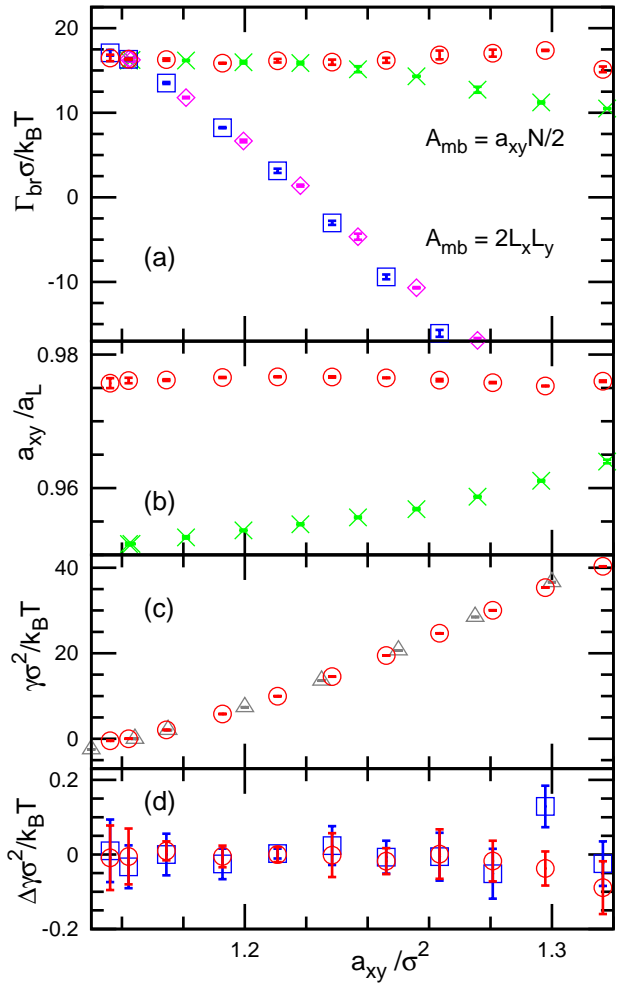}
\caption{\label{fig:ten}
Dependence of the junction line tension $\Gamma_{\rm {br}}$
and surface tension $\gamma$ on the membrane area $a_{xy}$ per molecule
for  $\varepsilon=8$, 
$k_{\rm {bend}}=8$, $k_{\rm {tilt}}=8$, and $C_{\rm {bd}}=0$.
(a) The line tension $\Gamma_{\rm {br}}$
calculated from Eq. (\ref{eq:pressure_y}) with $A_{\rm {mb}}=a_{xy}N/2$ ($\circ, \times$)
and $A_{\rm {mb}}=2_xLy$ ($\Box, \diamond$).
The symbols ($\times, \diamond$) and ($\circ, \Box$) represent data 
for $N=1536$ and $N=3072$, respectively. 
(b) The molecular area $a_{xy}$ for the bilayer
in the middle of the simulation box
is compared with
the  geometrically estimated area $a_{\rm L}=4L_xL_y/N$
for  $N=1536$  ($\times$) and $N=3072$ ($\circ$).
(c) The surface tension $\gamma$ 
calculated from the branched membranes ($\circ$) for $N=3072$ and the previous simulation 
of the planar bilayer membrane ($\triangle$) for $N=512$. 
(d) The differences in the values of $\gamma$ estimated from the three methods for $N=3072$.
The circles ($\circ$) and squares ($\Box$) represent $\gamma_{z} - \gamma_{x}$ and 
 $\gamma_{\rm {mid}}- \gamma_{x}$, respectively.
The tensions $\gamma_{x}$ and $\gamma_{z}$ are calculated from the pressure tensors $P_{xx}$
and $P_{zz}$, respectively. The tension $\gamma_{\rm {mid}}$ is calculated from
 the middle bilayer.
}
\end{figure}

\section{Measurement of line tension of branching junctions}
\label{sec:results}

To measure the stress of branched membranes,
we consider hexagonally arranged branches (see Fig. \ref{fig:hex_snap}).
These have an inverted hexagonal structure with a long cell length.
Three membranes  equiangularly connect with each other 
at the branching junction.
The periodic boundary along the $x$ direction 
is shifted as shown in Fig. \ref{fig:hex_snap}(a).
Normal periodic boundary conditions are employed
in the $y$ and $z$ directions.
Thus,  the periodic images are given by
 $(x_i+n_xL_x,y_i+n_yL_y,z_i+n_zL_z+n_xL_z/2)=(x_i,y_i,z_i)$,
where $n_x$, $n_y$, and $n_z$  are arbitrary integers
for a simulation box with side lengths $L_x$, $L_y$, and $L_z$.
Length $L_y$ is varied to change
the membrane area and surface tension.
The other side lengths are fixed 
as $L_x=\sqrt{3}L_z/2=28\sigma$ and $56\sigma$ 
 for $N=1536$ and $3072$, respectively.
When the membrane thickness is neglected,
the total membrane area in the simulation box
is $A_{\rm L}=2L_xL_y$,
since the area of one membrane is $2L_xL_y/3$ 
and three membranes exist in the box. 
The length of one branching junction is $L_y$,
and two junctions are in the box.
The origin $(x,y,z)=(0,0,0)$ is set at the center of the simulation box
and is kept at the center of the horizontal membrane.

The line tension of the branching junction
can be calculated from the diagonal components of the pressure tensor
\begin{equation}
\label{eq:pressure_tensor}
P_{\alpha\alpha} = \big(Nk_{\rm B}T - 
     \sum_{i} \alpha_{ij}\frac{\partial U}{\partial {\alpha}_{ij}} \big)/V,
\end{equation} 
where $\alpha \in \{x,y,z\}$ and $V=L_xL_yL_z$.
The stresses should be generated by
the line tension $\Gamma_{\rm {br}}$ of the  branching junction,
the surface tension $\gamma$, and the bulk pressure $P_{\rm {bulk}}$.
\begin{eqnarray}
\label{eq:pressure_x}
P_{xx}V &=&  P_{\rm {bulk}}V - \gamma L_xL_y, \\ \label{eq:pressure_y}
P_{yy}V &=&  P_{\rm {bulk}}V - \gamma A_{\rm {mb}} - 2\Gamma_{\rm {br}}L_y, \\ 
P_{zz}V &=&  P_{\rm {bulk}}V - \frac{\sqrt{3}}{2}\gamma L_yL_z.   \label{eq:pressure_z}
\end{eqnarray}
In the membranes with an angle of $\pm\pi/3$ to the $xy$ plane,
the surface tension yields stresses $\gamma/2$ and  $(\sqrt{3}/2)\gamma$
 in the $x$ and $z$ directions, respectively.
Since the membranes can be moved in the $x$ and $z$ directions,
these stresses are balanced {\it i.e.} $P_{xx}=P_{zz}$.
Indeed, Eqs. (\ref{eq:pressure_x}) and (\ref{eq:pressure_z}) give
the same value.
Since all membranes are parallel to the $y$ direction,
 the total membrane area $A_{\rm {mb}}$ in the simulation box
is used in Eq. (\ref{eq:pressure_y}).
Since the critical micelle concentration (CMC) is low
and no isolated molecules appear,
the bulk pressure $P_{\rm {bulk}}$ is negligibly small 
in our solvent-free simulations.

To estimate the line tension  $\Gamma_{\rm {br}}$,
the surface tension $\gamma$ has to be calculated first.
In the simulations, $\gamma$ can be estimated from $P_{xx}$ or $P_{zz}$. 
We called them $\gamma_x$ and $\gamma_z$, respectively.
We also calculated $\gamma$
from the middle of the bilayer membrane which is parallel to the $xy$ plane
using the definition for the planar interface,
$\gamma_{\rm {mid}}= \langle P_{zz} - (P_{xx} + P_{yy})/2\rangle L_z$
for $-L_x/12<x<L_x/12$.
The estimated values of $\gamma$ from these three methods agree very well
[see Fig. \ref{fig:ten}(d)].
We use the average value $\gamma = (\gamma_x + \gamma_z)/2$
as the surface tension to reduce  statistical errors.
Here, we only consider membranes with small or positive surface tensions,
{\it i.e.}, exclude the  buckled membranes under lateral compression,
since the surface tension becomes anisotropic in buckled membranes \cite{nogu11a}.

\begin{figure}
\includegraphics{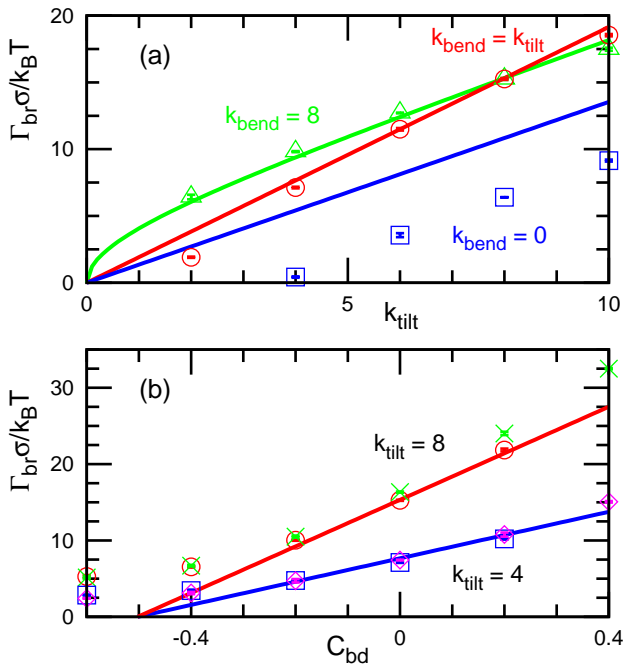}
\caption{\label{fig:bif}
Dependence of the line tension $\Gamma_{\rm {br}}$ of the branching junction
on (a) $k_{\rm {tilt}}$ and (b) $C_{\rm {bd}}$.
(a) The squares, circles, and triangles represent data 
for $k_{\rm {bend}}=0$, $k_{\rm {bend}}=k_{\rm {tilt}}$, and $k_{\rm {bend}}=8$, respectively,
at $\varepsilon=2$ and  $C_{\rm {bd}}=0$.
(b) The symbols ($\circ$, $\Box$) and ($\times, \diamond$)
represent data for  $\varepsilon=2$ and $8$, respectively,
at $k_{\rm {bend}}=k_{\rm {tilt}}=4$ or $8$.
The solid lines in (a) and (b) are obtained from Eq. (\ref{eq:Gbr_tilt}).
}
\end{figure}

Next, we consider the membrane area.
The projected area $a_{xy}$ per molecule is calculated 
from the middle of the membrane at $-L_x/12<x<L_x/12$
as $a_{xy}=(L_xL_y/6)/(N_{\rm {mid}}/2)$  where $N_{\rm {mid}}$ is the number of the 
molecules at $-L_x/12<x<L_x/12$.
The $a_{xy}$ dependence of $\gamma$ calculated from branched membranes
agrees with that obtained from planar membranes [see Fig. \ref{fig:ten}(c)].
This area $a_{xy}$ is slightly smaller than the area $a_{\rm L}=2L_xL_y/(N/2)$ calculated from the geometry. 
The decreases are $4$\% and $2$\%  for $N=1536$ and $3072$, respectively, as shown in Fig. \ref{fig:ten}(b).
These deviations are caused by the membrane thickness.
In Fig. \ref{fig:line_r}, the length of the membrane surface is shorter than the center line of the membrane.
Thus, the monolayer (leaflet) area of the bilayer membrane becomes  smaller than $a_{\rm L}$.

Finally, the line tension $\Gamma_{\rm {br}}$ of the  branching junction
is estimated from the pressure tensor and $a_{xy}$,
\begin{equation}
\label{eq:G_br}
\Gamma_{\rm {br}} = \Big\{ (P_{xx} + P_{zz})\frac{a_{xy}}{a_{\rm L}} - P_{yy} \Big\}\frac{V}{2L_y}.
\end{equation} 
In explicit solvent simulations ($P_{\rm {bulk}} \ne 0$), 
$\gamma$ or $P_{\rm {bulk}}$
should be calculated from a local region in the simulation box. 
When $\gamma_{\rm {mid}}$ is used, the line tension  is given by
\begin{equation}
\label{eq:G_br2}
\Gamma_{\rm {br}} = \Bigg\{ \Big(\frac{P_{xx} + P_{zz}}{2} - P_{yy}\Big)\frac{V}{2L_y}  +   \frac{\gamma_{\rm {mid}}L_x}{2}\Big(1 - \frac{2a_{xy}}{a_{\rm L}}\Big) \Bigg\},
\end{equation} 
for explicit solvent systems.
When $P_{\rm {bulk}}$ is estimated locally,
the line tension is given by
\begin{equation}
\label{eq:G_br3}
\Gamma_{\rm {br}} = \Bigg\{ (P_{xx} + P_{zz})\frac{a_{xy}}{a_{\rm L}} - P_{yy} + P_{\rm {bulk}}\Big( 1- \frac{2a_{xy}}{a_{\rm L}}  \Big) \Bigg\}\frac{V}{2L_y}.
\end{equation}

The tension $\Gamma_{\rm {br}}$ is almost independent of $\gamma$  [see Fig. \ref{fig:ten}(a)].
The estimated values of $\Gamma_{\rm {br}}$ for two system sizes $N=1536$ and $3072$
agree very well, so that neighboring junctions are  sufficiently far away
to avoid the finite size effects of the membranes.
If we naively use  $A_{\rm {mb}}=2L_xL_y$, the obtained values of $\Gamma_{\rm {br}}$
decrease with increasing $\gamma$, and even become negative [see Fig. \ref{fig:ten}(a)].
Thus, although the area difference between $a_{xy}$ and $a_{\rm L}$ is very small,
it has a large influence on the estimation of $\Gamma_{\rm {br}}$.
The monolayer membrane area $a_{xy}N/2$ should be employed.

We investigated the $\Gamma_{\rm {br}}$ dependence on the potential parameters in our membrane model,
as shown in Fig. \ref{fig:bif}.
The tension $\Gamma_{\rm {br}}$  is  roughly linearly dependent on $k_{\rm {tilt}}$
and $C_{\rm {bd}}$ while it is almost independent of $\varepsilon$.
These dependencies can be understood by the energy minimization in the tilt model
as explained in the next section.

\section{Tilt model}
\label{sec:tilt}

\begin{figure}
\includegraphics[width=8.5cm]{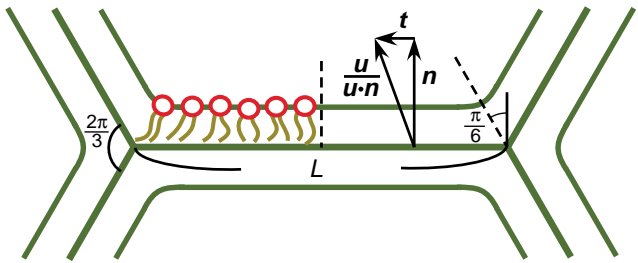}
\caption{\label{fig:line_r}
Schematic representation of the tilt model.
}
\end{figure}

We analyze the line tension of the branching junction and 
the stability of the inverted hexagonal $H_{II}$ phase
using the tilt model proposed by Hamm and Kozlov \cite{hamm98,hamm00}.
The orientation vector ${\bf u}$ of the lipid molecules can 
deviate from the normal vector ${\bf n}$ of the dividing surface of the bilayer.
The tilt vector is defined as
${\bf t}={\bf u}/{\bf u}\cdot{\bf n} - {\bf n}$,
which is parallel to the dividing surface (see Fig. \ref{fig:line_r}).
The tilt tensor is the derivative of the vector ${\bf t}$:
 $t^{\alpha}_{\beta}= \partial t_{\alpha}/\partial \beta$,
where $\alpha, \beta \in \{x,y\}$ for a membrane parallel to the $xy$ plane.
In the tilt model, the free energy of monolayers of the tensionless membranes is 
given by
\begin{equation}
F = \int \left[  \frac{\kappa_{\rm {mo}}}{2} ( t^{\alpha}_{\alpha}
 - C_0)^2 + \bar{\kappa}_{\rm {mo}}{\rm det}(t^{\alpha}_{\beta}) + \frac{\kappa_{\rm \theta}}{2} {\bf t}^2 \right] dA,
\label{eq:Ftilt}
\end{equation}
where  $t^{\alpha}_{\alpha}$ and det$(t^{\alpha}_{\beta})$ are the trace and determinant
of the tilt tensor, respectively.
The coefficients $\kappa_{\rm {mo}}$ and $\bar{\kappa}_{\rm {mo}}$
are the bending rigidity and saddle-splay modulus of the monolayer, respectively,
in the Helfrich-Canham bending energy \cite{safr94,helf73}.
The coefficient $\kappa_{\rm \theta}$ is the tilt modulus.

\begin{figure}
\includegraphics{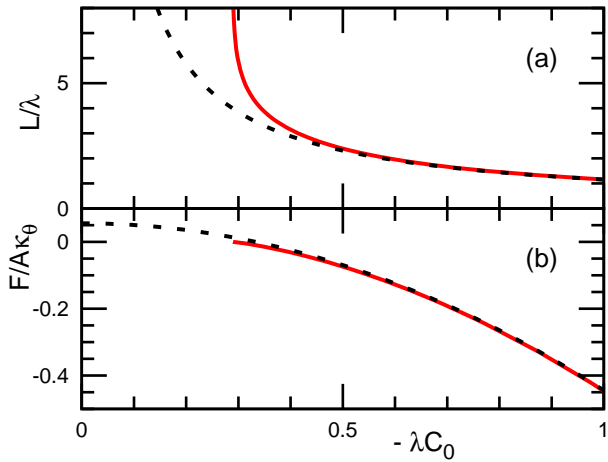}
\caption{\label{fig:l_c0}
The cell length $L$ (a) and free energy $F$ (b) of inverted hexagonal $H_{II}$ phase
with respect to the lamellar phase.
The solid and dashed lines are obtained from the tilt model 
with variation calculus [Eqs. (\ref{eq:Ftilt_mono})]
and  the linear approximation [Eq. (\ref{eq:Ftilt_hamm})], respectively.
}
\end{figure}

First, we consider the free energy of a horizontal monolayer in Fig. \ref{fig:line_r}.
Here, we use the same coordinates as the simulations [${\bf n}=(0,0,1)$].
Since the molecules tilt only in the $x$ direction, {\it i.e.}, $t_y=0$,
the free energy of the monolayer is expressed as
\begin{equation}
F_{\rm {hex}} =  2L_y \int_{0}^{L/2} \left[  \frac{\kappa_{\rm {mo}}}{2} ( t'_x(x)
 - C_0)^2  + \frac{\kappa_{\rm \theta}}{2} t_x(x)^2 \right] dx,
\label{eq:Ftilt}
\end{equation}
where $t'_x=\partial t_x/\partial x$.
The boundary conditions are $t_x(0)=0$ and $t_x(L/2)= -\tan(\pi/6)= -1/\sqrt{3}$,
where $L$ is the side length of the hexagonal cell, $L=2L_x/3$.
Hamm and Kozlov \cite{hamm98} obtained
\begin{equation}
\frac{F_{\rm {hex}}}{A} = \frac{\kappa_{\rm {mo}}}{2}\Big(\frac{2}{\sqrt{3}L} + C_0 \Big)^2 + \frac{\kappa_{\rm \theta}}{18},
\label{eq:Ftilt_hamm}
\end{equation}
from the assumption that $t'_x$ is constant.
Here, we use $t_x(x)$ which minimizes $F_{\rm {hex}}$, instead of the constant $t'_x$ assumption.
The calculus of variation gives
$t''_x(x) = (\kappa_{\rm \theta}/\kappa_{\rm {mo}})t_x(x)$.
Hence, the molecules are tilted by
\begin{equation}
t_x(x)= -\frac{\sinh(x/\lambda)}{\sqrt{3}\sinh(L/2\lambda)},
\label{eq:tilt_x}
\end{equation}
where $\lambda=\sqrt{\kappa_{\rm {mo}}/\kappa_{\rm \theta}}$ is the characteristic length.
Then the free energy is obtained as
\begin{equation}
\frac{F_{\rm {hex}}}{\kappa_{\rm {mo}} L_y} = \frac{\sinh(L/\lambda)}{3\lambda(\cosh(L/\lambda)-1)} + \frac{2C_0}{\sqrt{3}} + \frac{{C_0}^2L}{2}.
\label{eq:Ftilt_mono}
\end{equation}
When the distance $L$ between the junctions is sufficiently large ({\it i.e.} $L \gg \lambda$),
the first term on the R.H.S of Eq. (\ref{eq:Ftilt_mono}) becomes $1/3\lambda$.

The free energy $F_{\rm {lam}}$ of the lamellar phase is given by
$F_{\rm {lam}}/A= \kappa_{\rm {mo}} C_0^2/2$.
The line tension of the branching junction is the energy required to form a junction per unit length,
$\Gamma_{\rm {br}}L_y = 3(F_{\rm {hex}} - F_{\rm {lam}})$,
since one junction links six monolayers and one monolayer connects two junctions.
Thus, $\Gamma_{\rm {br}}$ is given by
\begin{equation}
\Gamma_{\rm {br}} = \sqrt{\kappa_{\rm {mo}} \kappa_{\rm \theta}}( 1 + 2\sqrt{3}\lambda C_0),
\label{eq:Gbr_tilt}
\end{equation}
for $L \gg \lambda$.

Next, we compare the line tension of the tilt model with our simulation results.
Molecular tilts are seen around the junctions in simulation snapshots [see Fig. \ref{fig:hex_snap}(b)].
In our membrane model,
the bending rigidity and spontaneous curvature of the monolayer
are given by
$\kappa_{\rm {mo}}/k_{\rm B}T \simeq 1.1(k_{\rm {bend}} + k_{\rm {tilt}})$ 
and 
 $C_0 \simeq \{\kappa_{\rm {bend}}/(\kappa_{\rm {bend}}+\kappa_{\rm {tilt}})\}C_{\rm {bd}}/\sigma$,
respectively, as described in Sec. \ref {sec:mempro}.
The tilt modulus $\kappa_{\rm \theta}$ is estimated as follows.
When thermal fluctuations are neglected,
the tilt energy $U_{\rm {tilt}}$ of a flat membrane with a constant tilt $t_x$ is given by
\begin{equation}
\label{eq:Uav_tilt1}
\frac{F_{\rm {tilt}}}{k_{\rm B}T} = \frac{k_{\rm{tilt}}}{4} {t_x}^2 N N_{\rm {nb}} \langle w_{\rm {cv}} \rangle, 
\end{equation}
where $N_{\rm {nb}}$ is the mean number of neighboring molecules.
Here the average  $\langle ({\bf t}\cdot\hat{\bf{r}} )^2 \rangle = {\bf t}^2/d$
is used for isotropic unit vectors $\hat{\bf{r}}$ with a spatial dimension $d=2$.
In the continuum limit, $F_{\rm {tilt}}$ should correspond 
to $F_{\rm {hex}}=\kappa_{\rm \theta}A{t_x}^2/2$ of the tilt model.
Thus, $\kappa_{\rm \theta}$ is given by
 $\kappa_{\rm \theta} \simeq 2 k_{\rm{tilt}}k_{\rm B}T/a_{xy}$
from $N_{\rm {nb}}=6$ and $\langle w_{\rm {cv}} \rangle=w_{\rm {cv}}(1.05\sigma) = 0.61$.
The line tension $\Gamma_{\rm {br}}$ calculated from Eq. (\ref{eq:Gbr_tilt}) 
reproduces the simulation results very well for most of the parameter ranges
in Fig. \ref{fig:bif}.
The tilt model overestimates $\Gamma_{\rm {br}}$ for $k_{\rm {bend}}=0$ in Fig. \ref{fig:bif}(a).
This deviation is likely caused by a too small characteristic length $\lambda$.
It is less than the molecular diameter $\sigma$:
 $\lambda \simeq 0.8\sigma$ and $1.1\sigma$
at  $k_{\rm {bend}}=0$ and $k_{\rm {bend}}=k_{\rm {tilt}}$, respectively,
for the tensionless membranes with $a_{xy} \simeq 1.2\sigma^2$.

Finally, we discuss the stability of the inverted hexagonal $H_{II}$ phase
using the tilt model.
For $- \lambda C_0 \leq  1/2\sqrt{3}$,
the free energy per area $F_{\rm {hex}}/A$
 monotonically decreases 
with increasing cell length $L$
[see  Eq. (\ref{eq:Ftilt_mono})].
In this region, the lamellar phase ($L\to \infty$) 
is more stable than the $H_{II}$ phase.
For $- \lambda C_0 > 1/2\sqrt{3}$, 
 $F_{\rm {hex}}/A$ is minimum at finite $L$
and the $H_{II}$ phase is more stable than  the lamellar phase.
In this region, the line tension $\Gamma_{\rm {br}}$ is negative,
and the $H_{II}$ hexagonal structure grows.
Figure \ref{fig:l_c0} shows $L$ and $(F_{\rm {hex}}-F_{\rm {lam}})/A$
obtained by the present method and the previous linear approximation \cite{hamm98}.
Under the linear approximation,
$(F_{\rm {hex}}-F_{\rm {lam}})/A= \kappa_{\rm \theta}/18 - \kappa_{\rm {mo}} C_0^2/2$ with $L= - 2/\sqrt{3}C_0$,
so that the $H_{II}$ phase has a lower energy for $- \lambda C_0 > 1/3$. 
As $- \lambda C_0$ increases, the two lines in Fig. \ref{fig:l_c0} approach each other.
Thus, the linear approximation works well in the $H_{II}$ phase at $- \lambda C_0 > 0.5$,
while not for $- \lambda C_0 \lesssim  1/2\sqrt{3}$.

\begin{figure}
\includegraphics{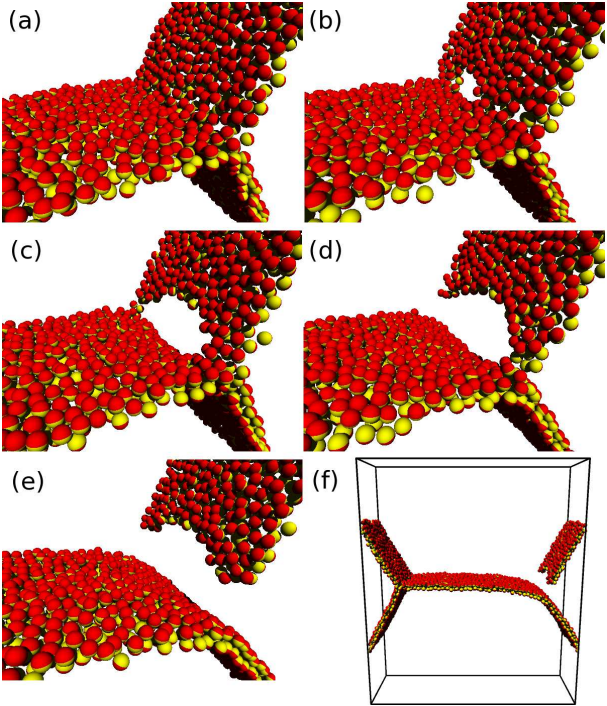}
\caption{\label{fig:rup_snap}
Sequential snapshots of the membrane rupture
for $N=3072$, $\varepsilon=2$, 
$k_{\rm {bend}}=8$, $k_{\rm {tilt}}=8$, and $C_{\rm {bd}}=0$.
(a) $t/\tau_0=2555$. (b) $t/\tau_0=2565$. (c) $t/\tau_0=2570$. (d) $t/\tau_0=2575$. 
(e), (f) $t/\tau_0=2580$. 
}
\end{figure}

\begin{figure}
\includegraphics{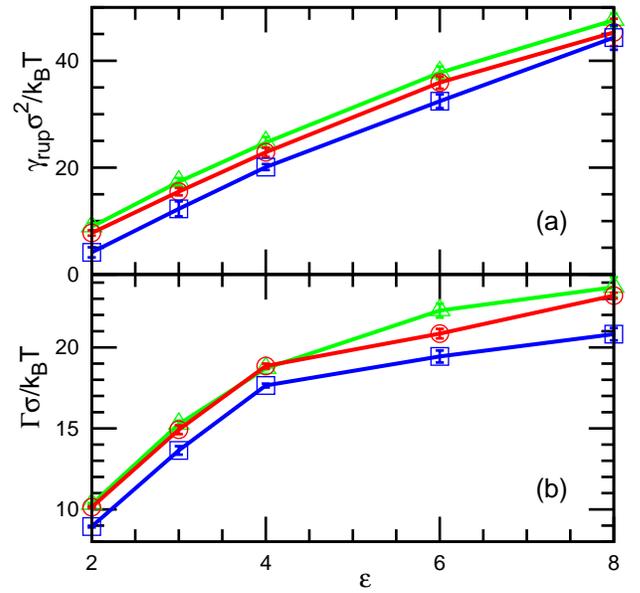}
\caption{\label{fig:cut}
Parameter $\varepsilon$ dependence of
(a) surface tension $\gamma_{\rm {cut}}$ of the membrane rupture
and (b) the line tension of the membrane edge
for $k_{\rm {bend}}=k_{\rm {tilt}}=8$.
The triangles, circles, and squares represent data for
 $C_{\rm {bd}}= -0.2$, $0$, and $0.2$, respectively.
}
\end{figure}

\section{Dynamics}
\label{sec:dynamic}

\subsection{Rupture}

A membrane rupture occurs under sufficiently high surface tension $\gamma$.
We investigated the rupture of branched membranes as
 $L_y$ is increased slowly at a rate $dL_y/dt=0.0002\sigma/\tau$.
We checked that this rate is sufficiently small.
No difference is detected when a rate twice as fast, {\it i.e.}, $dL_y/dt=0.0004\sigma/\tau$, is used.
Figure \ref{fig:rup_snap} shows the rupture of the  branched membranes.
A pore opens  on the side of the branching junction
and then expands along the junction.
Since the junction is less stable than the middle of the bilayer membranes,
the rupture always occurs on  the side of the branching junction.

The surface tension $\gamma_{\rm {rup}}$ at the rupture
is estimated from the extrapolation of the $L_{y}$-$\gamma$ line to the rupture point.
As $\varepsilon$ increases,  $\gamma_{\rm {rup}}$ increases linearly as shown in Fig. \ref{fig:cut}(a),
although $\Gamma_{\rm {br}}$ is independent of  $\varepsilon$.
The line tension $\Gamma$ of the membrane edge also increases
but it is not linear with respect to $\varepsilon$  [see Fig. \ref{fig:cut}(b)].

After the membrane rupture,
a branched membrane becomes
a line of a membrane edge and a bent membrane.
The free energy of the former and latter states of the tensionless membranes
are given by $F_{{\rm {br}}}/L_y= \Gamma_{\rm {br}}$ and $F_{{\rm {rup}}}/L_y= \Gamma + (\pi/6)\kappa/R$, respectively,
where $R \sim 5 \sigma$ is the curvature radius of the bent membrane.
Thus, the branched membrane is kinetically stable for the rupture when $\Gamma_{\rm {br}}<\Gamma$
for $\gamma=0$.
Furthermore, even when $F_{{\rm {br}}}> F_{{\rm {rup}}}$,
the branched membrane can remain in a metastable state
 to overcome the free energy barrier to rupture.
In the present spin molecular model, 
$\Gamma_{\rm {br}}$ and $\Gamma$ can be varied separately
by $k_{\rm {tilt}}$ or $k_{\rm {bend}}$ and $\varepsilon$, respectively.

\begin{figure}
\includegraphics[width=8.cm]{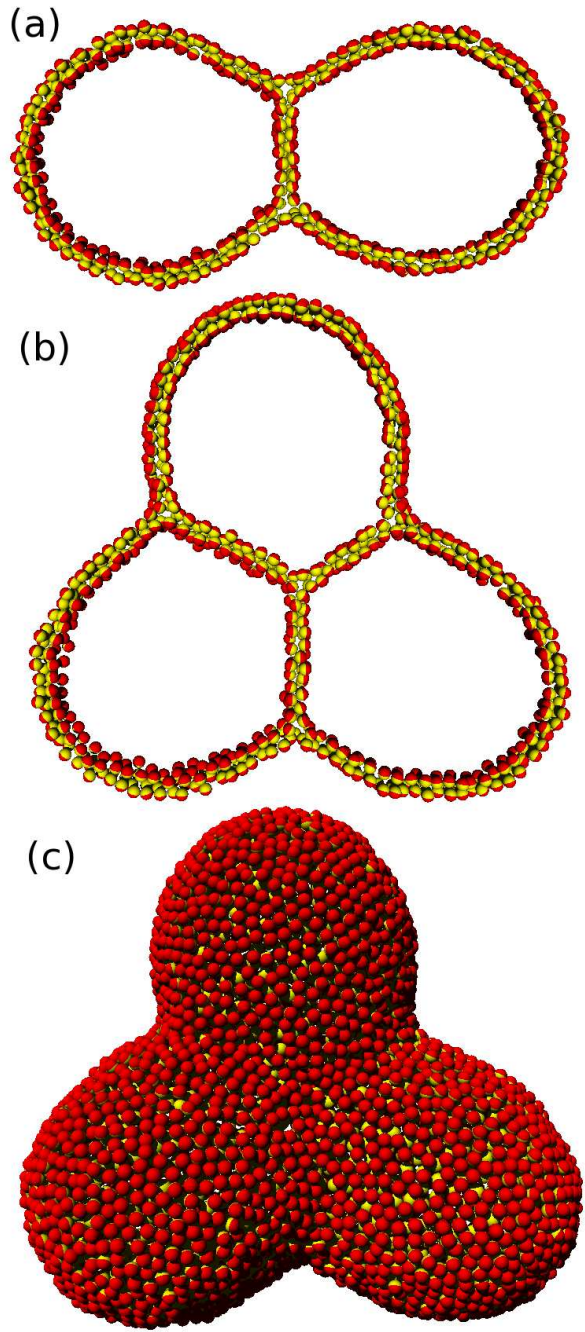}
\caption{\label{fig:tri_snap}
Branched structures in partially fused vesicles
for $\varepsilon=8$, $k_{\rm {bend}}=8$, $k_{\rm {tilt}}=8$, and $C_{\rm {bd}}=0$.
Sliced snapshots are shown in (a) two and (b) three hemifused vesicles
for $N=4000$ and $N=6000$, respectively.
(c) All molecules are shown for the snapshot in (b).
}
\end{figure}

\subsection{Fusion Intermediate}

The branching junctions appear in the hemifusion diaphragm (HD) intermediate 
in  membrane fusion
[see Fig. \ref{fig:tri_snap}(a)].
The HD intermediate was proposed in the original stalk model of membrane fusion~\cite{mark84}
and  was observed in experiments \cite{niko11} and
in some of the molecular simulations \cite{nogu02c,marr03,smei06,gao08}.
Here we focus on the structure of the HD intermediate
and the detail of the fusion dynamics of the present model will be described somewhere else.
Following the procedure in  our previous fusion simulations \cite{nogu02c},
a spherical particle with a radius of $4\sigma$ is placed at the center of each vesicle
and constant external forces are applied to the spherical particles.
The membranes pinched by the particles are fused either directly or via the HD intermediate.
The fusion dynamics are qualitatively similar to our previous simulations \cite{nogu02c}.
When the HD is formed,
we removed the spherical particles and then relaxed the vesicles to a stable structure.
The resultant structures are shown in Fig. \ref{fig:tri_snap}.

The HD of two vesicles has a circular branching junction. 
As the radius $R_{\rm {br}}$ of the branching junction increases,
the bending energy is reduced because the flat membrane area also increases.
However, the energy of the junction increases as $2\pi R_{\rm {br}}\Gamma_{\rm {br}}$.
Thus, the radius is determined by the balance of these energies.
Figures \ref{fig:tri_snap} (b) and (c) show the HD of three hemifused vesicles.
Four lines of the branching junction meet at the centers of the front and back membranes.
These vesicle shapes resemble the shape of soap bubbles.
In the bubbles, competition between surface tension and pressure maintains their shapes,
while in the vesicles, the bending energy and line tension play these roles.

\section{Summary}
\label{sec:sum}

We have studied the branching of  bilayer membranes.
The free energy of the branching junction
can be treated as line tension 
when the distance between the junctions is sufficiently large.
The line tension $\Gamma_{\rm {br}}$ of the junction is calculated
from the pressure tensor in solvent-free molecular simulations.
The simulation results agree well with the prediction of the
tilt model.

The stability of the branching membranes depends on the boundary conditions.
High surface tension induces  membrane rupture on the side of the branching junction.
In the hemifusion diaphragm intermediate,
the branching junctions can be formed by the balance of the bending energy and line tension.
Several kinds of proteins are known to modify membrane curvatures 
in living cells \cite{baum10,phil09,zimm06}.
The branched structures induced by the absorption of 
a colloid \cite{nogu02a}, DNA \cite{fara09}, and
 peptide \cite{kawa11} have been reported in molecular simulations.
The branching structures of the biomembranes may be stabilized or controlled by the
absorption or insertion of proteins to the membranes in living cells.

The self-assembled structures of amphiphilic molecules can be clarified from
the packing parameter \cite{isra11} of amphiphilic molecules 
or the spontaneous curvature of the monolayer.
However, these quantities are not easy to measure.
Alternatively, the line tension of the branching junction may be suitable to quantify
the tendency of the amphiphilic molecules to form inverted structures.

\begin{acknowledgments}
This study is partially supported by a Grant-in-Aid for Scientific Research 
on Priority Area ``Molecular Science of Fluctuations toward Biological Functions'' from
the Ministry of Education, Culture, Sports, Science, and Technology of Japan.
\end{acknowledgments}

\end{document}